\documentclass[pre, floatfix, twocolumn]{revtex4-1}
\usepackage[dvips]{graphicx}
\usepackage{txfonts}

\begin{document}

\title{Exact solution of the hydrodynamical Riemann problem with nonzero tangential velocities and the ultrarelativistic equation of state}

\author{Patryk Mach}
\affiliation{M. Smoluchowski Institute of Physics, Jagiellonian University, Reymonta 4, 30-059 Krak\'{o}w, Poland}
\author{Ma\l gorzata Pi\c{e}tka}
\affiliation{M. Smoluchowski Institute of Physics, Jagiellonian University, Reymonta 4, 30-059 Krak\'{o}w, Poland}

\begin{abstract}
We give a solution of the Riemann problem in relativistic hydrodynamics in the case of ultrarelativistic equation of state and nonvanishing components of the velocity tangent to the initial discontinuity. Simplicity of the ultra-relativistic equation of state (the pressure being directly proportional to the energy density) allows us to express this solution in analytical terms. The result can be used both to construct and test numerical schemes for relativistic Euler equations in $(3 + 1)$ dimensions.
\end{abstract}

\maketitle

\section{Introduction}

Solutions of the Riemann problem in the relativistic hydrodynamics are of crucial importance for the construction of modern numerical schemes designed to solve relativistic Euler equations. In most such schemes, it is the Riemann solver (usually an approximate one) that is responsible for the accuracy of the method and a proper resolution of possible shock waves \cite{living_review}. Moreover, quite recently Aloy and Rezzolla applied the analysis of solutions of the relativistic hydrodynamical Riemann problem to explain a boosting mechanism occurring in astrophysical jets, proving that the importance of such solutions is not merely academic \cite{aloy_rezzolla}.

Here by the Riemann problem we understand a Cauchy problem for a hyperbolic system of partial differential equations, where initial data consist of two constant states separated by a discontinuity in a form of a plane surface. In the case of hydrodynamics, such an initial discontinuity decays, giving rise to three possible elementary waves: a shock wave, a rarefaction wave, and the so-called contact discontinuity. The solution of the Riemann problem is thus a non-trivial one, and its precise form requires investigation.

The relativistic shock-tube problem, i.e., a Riemann problem with zero initial velocities, was investigated by Thompson in \cite{thompson}. Later, the Riemann problem in one spatial dimension was solved for the ultra-relativistic equation of state by Smoller and Temple \cite{smoller_temple} and for the perfect gas equation of state by Mart\'{\i} and M\"{u}ller \cite{marti_mueller}. The latter work was generalized by Pons, Mart\'{\i} and M\"{u}ller to the case, in which the fluid is allowed to move in the direction tangent to the discontinuity \cite{pons_marti_mueller}, but, due to the complexity of equations, the solution had to be computed numerically. Then, in \cite{rezzolla_zanotti, rezzolla_zanotti_lett, rezzolla_zanotti_pons} Rezzolla, Zanotti and Pons introduced a particularly convenient way of classifying the solutions, based on the relative velocity between both Riemann states.

An important progress was made by Giacomazzo and Rezzolla, who analyzed the Riemann problem in relativistic magnetohydrodynamics \cite{giacomazzo_rezzolla}. A numerical code written for this analysis was then used to obtain some test solutions of the Riemann problem for equations of state other than that of perfect gas \cite{meliani_keppens_giacomazzo}.

In this paper we present an analytic solution for the Riemann problem with non-zero velocities tangent to the initial discontinuity and the ultra-relativistic equation of state. This common equation of state is exceptional in the sense that it cannot be expressed in terms of the baryonic (rest mass) density and the specific internal energy, but it relates the pressure directly to the energy density. This fact prevented us from a straightforward application of the existing numerical schemes solving relativistic Riemann problem. As a benefit we got a solution that can be expressed almost entirely in analytical terms.

Solutions of the Riemann problem in which tangential velocities do not vanish can be used to construct general numerical schemes that solve equations of hydrodynamics in all three spatial dimensions. A solver of this kind has been implemented using the solution discussed in \cite{pons_marti_mueller}, although it is not given in analytical terms, and, in order to obtain such a solution, one has to integrate a certain ordinary differential equation numerically. In the case presented here, the appropriate ordinary differential equation was solved analytically, so the implementation of the exact Riemann solver is straightforward.

We should also note that the effects caused by the presence of the tangential velocities in the Riemann problem are purely relativistic. In Newtonian hydrodynamics they do not influence the behavior of the solution in the direction normal to the discontinuity. Thus, in order to extend a given one-dimensional solution to the case with non-zero tangential velocities, it is only required to compute the values of those velocities in the intermediate states. In relativistic hydrodynamics all velocities couple together through Lorentz factors, and the presence of tangential velocities changes the solution quantitatively.

Throughout this work we will assume that the reader has a basic knowledge of the Riemann problem for general sets of nonlinear hyperbolic equations (a good introduction can be found in \cite{evans}). In Secs. II and III we will review basic equations constituting our problem. Afterward, in Secs. IV--VI we will discuss the structures of rarefaction waves, shock waves and contact discontinuities respectively. Next, in Sec. VII the solutions of the Riemann problem will be presented, and in Sec. VIII we will compare them to the solutions obtained for the perfect gas equation of state. A summary of the paper will be given in Sec. IX.


\section{Relativistic Euler equations and the equation of state}

The equations expressing the conservation of the energy and momentum in relativistic hydrodynamics are usually written in the following compact form
\begin{equation}
\label{energy_momentum_cons}
\partial_\mu T^{\mu \nu} = 0,
\end{equation}
where  the energy-momentum tensor is that of perfect fluid, namely
\begin{equation}
T^{\mu \nu} = (\rho + p)u^\mu u^\nu + p \eta^{\mu \nu}.
\end{equation}
Here $\rho$ denotes the energy density, $p$ is the pressure, $u^\mu$ are the components of the four-velocity of the fluid, and $\eta^{\mu \nu} = \mathrm{diag}(-1,+1,+1,+1)$ is the metric tensor of the Minkowski space-time. Throughout this paper Greek indices will refer to space-time dimensions ($\mu = 0, 1, 2, 3$), while Latin ones will be reserved for spatial dimensions ($i = 1, 2, 3$). We will also work in Cartesian coordinates, where $x^\mu = (t,x,y,z)$.

In order to solve the Riemann problem for Eqs.~(\ref{energy_momentum_cons}) it is convenient to rewrite them in the form where the derivatives with respect to time and spatial coordinates are separated explicitly. To this end, we introduce the Lorentz factor $W=u^0$ and components of the three-velocity $v^i = u^i/W$. Due to the normalization of the four-velocity $\eta_{\mu \nu} u^\mu u^\nu = -1$, the Lorentz factor can be written as $W = 1/\sqrt{1 - v_i v^i}$. In the above terms Eqs.~(\ref{energy_momentum_cons}) can be expressed as
\begin{equation}
\partial_t \mathbf U + \partial_i \mathbf F^i = 0,
\end{equation}
where
\begin{eqnarray}
\mathbf U & = & \left( (\rho + p) W^2 - p, (\rho + p) W^2 v^1, (\rho + p) W^2 v^2, \right. \nonumber \\
& & \left. (\rho + p) W^2 v^3 \right)^T,
\end{eqnarray}
and
\begin{eqnarray}
\mathbf F^i & = & \left( (\rho + p) W^2 v^i, (\rho + p) W^2 v^i v^1 + \delta^{i 1} p, \right. \nonumber \\
& & \left. (\rho + p) W^2 v^i v^2 + \delta^{i 2}p, (\rho + p) W^2 v^i v^3 + \delta^{i 3} p  \right)^T.
\end{eqnarray}
Here $\delta^{ij}$ denotes the Kronecker's delta.

By the ultra-relativistic equation of state we understand a relation $p = c^2_s \rho$, where $c_s \in (0,1)$ is a constant playing the role of the local speed of sound  (for a photon gas or a gas of neutrinos $c^2_s = 1/3$). This form of equation of state is commonly used in cosmology; this is also the equation of state assumed in \cite{smoller_temple}. For the ultra-relativistic equation of state (a barotropic equation of state of the form $p = p(\rho)$, in general) Eqs.~(\ref{energy_momentum_cons}) constitute a complete set of equations of hydrodynamics.

On the other hand, the perfect gas equation of state, exploited in most of numerical simulations in relativistic hydrodynamics and in \cite{marti_mueller, pons_marti_mueller}, has the form $p = (\gamma - 1)n\epsilon$, where $\gamma$ is a constant, $n$ is the so-called baryonic (or rest mass) density and $\epsilon$ denotes the specific internal energy. The baryonic density is assumed to be a function satisfying the following continuity equation
\begin{equation}
\label{baryonic_density_cons}
\partial_\mu(nu^\mu) = 0,
\end{equation}
and the specific internal energy is defined as $\epsilon = (\rho - n)/n$. Thus, for the perfect gas equation of state the equations of hydrodynamics consist of Eqs.~(\ref{energy_momentum_cons}) and (\ref{baryonic_density_cons}).

For some physical situations the baryonic density is much smaller than the energy density and $\rho = n + n \epsilon \approx n \epsilon$. In this case relations $p = c^2_s \rho$ and $p = (\gamma - 1)n \epsilon$ should be equivalent, provided that $\gamma - 1 = c^2_s$. The equations of hydrodynamics suitable for these two equations of state are, however, different and the solutions can differ even qualitatively (there is, for instance, no contact discontinuity for the ultra-relativistic equation of state and no tangential velocities in the Riemann problem, as the pressure is directly proportional to the energy density, and such a discontinuity is present in an analogous solution for the perfect gas equation of state, where the same pressure can correspond to different values of baryonic density). A careful inspection of solutions of the Riemann problem in both cases shows that they tend to each other in a suitable sense. It should, however, be noted that in our case of ultrarelativistic equation of state the solution of the appropriate Riemann problem can be found analytically, whereas it was not possible for the case of perfect gas equation of state \cite{pons_marti_mueller}.

In this paper, we specialize to the ultrarelativistic equation of state, although many results are more general, valid for barotropic equations of state $p = p(\rho)$.


\section{Riemann problem}

Without loss of generality, we will assume that the initial discontinuity is perpendicular to the $x$ axis. Thus, neglecting the derivatives with respect to $y$ and $z$, we can write equations for the Riemann problem as
\begin{eqnarray}
\partial_t \left( (\rho + p) W^2 - p \right) + \partial_x \left( (\rho + p) W^2 v^x \right) & = & 0, \label{set_a} \\
\partial_t \left( (\rho + p) W^2 v^x \right) + \partial_x \left( (\rho + p) W^2 (v^x)^2 + p \right) & = & 0, \label{set_b} \\
\partial_t \left( (\rho + p) W^2 v^y \right) + \partial_x \left( (\rho + p) W^2 v^x v^y \right) & = & 0, \label{set_c} \\
\partial_t \left( (\rho + p) W^2 v^z \right) + \partial_x \left( (\rho + p) W^2 v^x v^z \right) & = & 0. \label{set_d}
\end{eqnarray}

The structure of solutions of the relativistic Riemann problem is exactly the same as in the corresponding Newtonian case, and it is, in fact, shared by general sets of hyperbolic conservation laws (cf.~\cite{evans}). Let the initial discontinuity be located at $x = 0$, and let $L$ and $R$ refer to the left and right Riemann states, that is data for $x < 0$ and $x > 0$, respectively. The form of Eqs.~(\ref{set_a})--(\ref{set_d}) and the symmetry of the initial data suggest a self-similar solution depending on $x$ and $t$ through $\xi = x/t$ only. The initial state $LR$ decays into three possible elementary self-similar waves separated by some constant states. A smooth elementary wave, the so called rarefaction wave will be further denoted by $\mathcal R_{\rightarrow(\leftarrow)}$, where the subscript arrows refer to the direction from which particles of the fluid enter the wave. The other two elementary waves are discontinuities: a shock wave, denoted by $\mathcal S_{\rightarrow(\leftarrow)}$, and a contact discontinuity $\mathcal C$. We will also use the symbol $\mathcal W_{\rightarrow(\leftarrow)}$ to denote a shock wave $\mathcal S_{\rightarrow(\leftarrow)}$ or a rarefaction wave $\mathcal R_{\rightarrow(\leftarrow)}$, when the actual character of the wave is not important.

The decay of the initial state $LR$ can be symbolically written as
\begin{equation}
LR \to L \mathcal W_\leftarrow L_\ast \mathcal C R_\ast \mathcal W_\rightarrow R,
\end{equation}
which corresponds to four different cases with $\mathcal W_{\rightarrow (\leftarrow)} = \mathcal S_{\rightarrow (\leftarrow)}$ or $\mathcal W_{\rightarrow (\leftarrow)} = \mathcal R_{\rightarrow (\leftarrow)}$.

The distinction between a contact discontinuity $\mathcal C$ and a shock wave $\mathcal S$ is based on the behavior of the pressure and the normal velocity ($v^x$ in our case) across the discontinuity. They are assumed to be continuous at the contact discontinuity and exhibit a jump at the shock wave. Since for the ultrarelativistic equation of state the pressure is directly proportional to the energy density, the only quantities that can be discontinuous across a contact discontinuity are the tangential components of the velocity ($v^y$ and $v^z$). This also means that the pressure, the normal velocity $v^x$, and, in case of ultra-relativistic equation of state, the energy density are the same in both intermediate states $L_\ast$ and $R_\ast$. 

The strategy of finding of the solution of the Riemann problem can be now summarized as follows. We start by considering a left moving wave $\mathcal W_\leftarrow$, and obtain the relation between the energy density $\rho_{L_\ast}$ and the velocity $v^x_{L_\ast}$ in the region behind such a wave. Next, we repeat the same calculations for the right moving wave $\mathcal W_\rightarrow$, to obtain an analogous relation between the energy density $\rho_{R_\ast}$ and the velocity $v^x_{R_\ast}$. Since the energy density and the normal velocity are the same in both intermediate states, they can be computed from the equation $\rho_{L_\ast} (v^x_{L_\ast}) = \rho_{R_\ast} (v^x_{R_\ast})$. The solution to this equation also identifies the actual character of both waves $\mathcal W_\rightarrow$ and $\mathcal W_\leftarrow$ (the so-called wave pattern). We will discuss the actual forms of the relation between the energy density and the normal velocity for all kinds of simple waves in the forthcoming sections.


\section{Rarefaction wave}

Let us first consider a rarefaction wave, that is, a smooth self-similar solution depending on $t$ and $x$ through $\xi = x/t$ only. In this case Eqs.~(\ref{set_a})--(\ref{set_d}) reduce to
\begin{eqnarray}
\label{set2_1}
\xi \frac{d}{d \xi} \left( (\rho + p) W^2 - p \right) & = & \frac{d}{d \xi} \left( (\rho + p) W^2 v^x \right), \\
\label{set2_2}
\xi \frac{d}{d \xi} \left( (\rho + p) W^2 v^x \right) & = & \frac{d}{d \xi} \left( (\rho + p) W^2 (v^x)^2 + p \right), \\
\label{set2_3}
\xi \frac{d}{d \xi} \left( (\rho + p) W^2 v^y \right) & = & \frac{d}{d \xi} \left( (\rho + p) W^2 v^x v^y \right), \\
\label{set2_4}
\xi \frac{d}{d \xi} \left( (\rho + p) W^2 v^z \right) & = & \frac{d}{d \xi} \left( (\rho + p) W^2 v^x v^z \right).
\end{eqnarray}

Nontrivial solutions of these equations exist only if the Wronskian of the above set of equations vanishes, i.e., when $\xi$ are the eigenvalues of the Jacobian $\partial \mathbf F^x / \partial \mathbf U$. Such eigenvalues can be easily found by exploiting the following observation. Let $\mathbf \Sigma = (\rho, v^x, v^y, v^z)$, $\mathcal A = \partial \mathbf U / \partial \mathbf \Sigma$ and $\mathcal B = \partial \mathbf F^x / \partial \mathbf \Sigma$. For a barotropic equation of state where $p = p(\rho)$ and $c^2_s = dp/d\rho$, the determinant of $\mathcal A$ reads
\begin{equation}
\mathrm{det} \mathcal A = W^8 (\rho + p )^3 \left( 1 - v_i v^i c^2_s \right).
\end{equation}
Since it is positive for $c^2_s \in (0,1)$, the matrix $\mathcal A$ is invertible and $\partial \mathbf F^x / \partial \mathbf U = \mathcal B \mathcal A^{-1}$. Consider
\begin{equation}
\mathrm{det} \left( \mathcal B \mathcal A^{-1} - \xi \mathcal{I} \right) \mathrm{det} \mathcal A = \mathrm{det} \left( \mathcal B - \xi \mathcal A \right),
\end{equation}
where $\mathcal I$ denotes the identity matrix. Since $\mathrm{det} \mathcal A \neq 0$, it is clear that the values of $\xi$ satisfying $\mathrm{det} \left( \mathcal B - \xi \mathcal A \right) = 0$ are the eigenvalues of the Jacobian $\partial \mathbf F^x / \partial \mathbf U$. They can be easily computed to yield
\[ \xi_0 = v^x, \]
\begin{equation}
\label{eigenvalues}
\xi_\pm = \frac{v^x (1 - c^2_s) \pm c_s \sqrt{\left( 1 - v_i v^i \right) \left( 1 - v_i v^i c^2_s - (v^x)^2 (1 - c^2_s) \right)}}{1 - v_i v^i c^2_s}.
\end{equation}
The eigenvalue $\xi_0$ is twofold degenerate. The expression for $\xi_\pm$ can be also written as
\begin{equation}
\xi_\pm = \frac{v^x \pm A}{1 \pm v^x A},
\end{equation}
with $A^{-2} = 1 + W^2 \left( 1 - (v^x)^2 \right)(1 - c^2_s)/c^2_s$, where we recognize the relativistic composition law for velocities. The values $\xi_+$ and $\xi_-$ correspond, respectively, to the signals propagating to the right (towards larger values of $x$) and to the left with respect to the local flow of gas. For $v^y = v^z = 0$ we have $A = c_s$ so that $c_s$ can be identified with the local speed of sound. It is also worth pointing out that the same expressions can be obtained in a case where the pressure depends on the baryonic density $n$ and the specific internal energy $\epsilon$ with, $c^2_s = d p / d \rho$ being replaced by
\begin{equation}
c^2_s = \frac{1}{h} \left( \left(\frac{\partial p}{\partial n} \right)_{\epsilon} +  \frac{p}{n^2} \left( \frac{\partial p}{\partial \epsilon} \right)_n \right),
\end{equation}
where $h = 1 + \epsilon + p/n$ is the specific enthalpy. In the latter case, however, we are dealing with 5 instead of 4 equations, and the value $\xi_0 = v^x$ is threefold degenerate \cite{pons_marti_mueller}.

It can be deduced from Eqs.~(\ref{set_a})--(\ref{set_d}) that, as long as we are interested in a smooth solution, the entropy density $s$ defined by
\begin{equation}
\label{artif_n}
s = s_1 \mathrm{exp} \int_{\rho_1}^\rho \frac{d \rho^\prime}{\rho^\prime + p \left( \rho^\prime \right)},
\end{equation}
where $s_1$ and $\rho_1$ are constants, satisfies the equation
\begin{equation}
\label{s_cons}
\partial_t \left( s W \right) + \partial_x\left( s W v^x \right) = 0,
\end{equation}
and thus
\begin{equation}
\label{help}
\xi \frac{d}{d \xi} \left( s W \right) = \frac{d}{d \xi} \left( s W v^x \right).
\end{equation}
It should be pointed out that the entropy density given by Eq.~(\ref{artif_n}) is not conserved for discontinuous solutions, that is the Rankine--Hugoniot conditions following from Eq.~(\ref{s_cons}) are not satisfied. In the case of ultrarelativistic equation of state the integral appearing in Eq.~(\ref{artif_n}) can be evaluated to yield $s = C \rho^{1/(1 + c^2_s)}$, where $C$ is a constant. Combining Eq.~(\ref{set2_3}) and (\ref{help}) gives
\begin{equation}
(\xi - v^x) \frac{d}{d\xi} \left( \rho^\kappa W v^y \right) = 0,
\end{equation}
where $\kappa = c^2_s/(1 + c^2_s)$. A similar result holds for Eq.~(\ref{set2_4}) and $v^z$, so that for $\xi \neq v^x$ we obtain
\begin{equation}
\rho^\kappa W v^y = \mathrm{const}, \;\;\; \rho^\kappa W v^z = \mathrm{const}.
\end{equation}

Let us introduce the tangential velocity $v^t$ as $v^t = \sqrt{(v^y)^2 + (v^z)^2}$. It follows that $v^t = a W^{-1} \rho^{-\kappa}$, where $a$ denotes a constant. Thus, from the definition of the Lorentz factor we have
\begin{equation}
\label{R_tilde}
W^2 \left( 1 - (v^x)^2 \right) = 1 + a^2 \rho^{-2\kappa} \equiv \tilde R(\rho).
\end{equation}
A little longer calculation shows that
\begin{equation}
\left( \xi - v^x \right) W^2 d v^x = \left( 1 - \xi v^x \right) d \ln \rho^\kappa.
\end{equation}
Inserting the expression for $\xi = \xi_\pm$ into this equation and performing some algebra, one can arrive at the following relation
\begin{equation}
\pm \frac{d v^x}{1 - (v^x)^2} = \frac{\sqrt{\tilde R + c^2_s(1 - \tilde R)}}{\tilde R c_s} d \ln \rho^\kappa.
\end{equation}
Both sides of this equation can be integrated, but the precise form of the result depends on the value of the constant $a$. For $a = 0$ (no tangential velocities) we obtain
\begin{equation}
\left( \frac{1 + v^x}{1 - v^x} \right)^{\pm \frac{1}{2}} = C_1 \rho^{( \kappa / c_s)}.
\end{equation}
For non-zero tangential velocities one gets
\begin{eqnarray}
\left( \frac{1 + v^x}{1 - v^x} \right)^{\pm 1} & = & C_2 \left( \frac{1 + \sqrt{1 + (1 - c^2_s) a^2 \rho^{-2\kappa}}}{1 - \sqrt{1 + (1 - c^2_s) a^2 \rho^{-2\kappa}}} \right)^{(1 / c_s)} \nonumber \\
& & \times \frac{c_s - \sqrt{1 + (1 - c^2_s) a^2 \rho^{-2\kappa}}}{c_s + \sqrt{1 + (1 - c^2_s) a^2 \rho^{-2\kappa}}}.
\end{eqnarray}
Knowing the state ahead the rarefaction wave we can thus compute the appropriate integration constant ($C_1$ or $C_2$) and obtain the solution in the region behind the front of the wave. The characteristics corresponding to this solution, treated as curves in the $(t,x)$ space, form a ``rarefaction fan,'' in which each characteristic correspond to a different value of $\xi_+$ (for the right moving wave) or $\xi_-$ (for the left moving one). For $\xi = \xi_0 = v^x$ we obtain $d \rho / d \xi = d v^x / d \xi = 0$, so that the ``fan'' of characteristics originating at the discontinuity has a ``zero opening angle.'' Remarkably, Eqs.~(\ref{set2_1})--(\ref{set2_4}) give no conditions for $v^x$ and $v^y$ in this case. This corresponds to the contact discontinuity which will be treated later in this paper.


\section{Shock wave}

Rankine--Hugoniot conditions for Eqs.~(\ref{energy_momentum_cons}) can be written as
\begin{equation}
\left[ \left[ T^{\mu\nu} \right] \right] n_\mu = 0,
\end{equation}
where $n^\mu$ is the unit vector normal to the surface of discontinuity and $\left[ \left[ f \right] \right]$ represents the jump of a given quantity $f$ at the discontinuity. Since we are interested in establishing the state behind the wave basing on the state in front of it (the left or the right state in the Riemann problem depending on the direction in which the wave propagates), we will adopt a notation in which values referring to the state in front of the shock wave are denoted with a bar while unaltered symbols are reserved for the values behind the discontinuity. In such a notation, a jump of a quantity $f$ reads $\left[ \left[ f \right] \right] = f - \bar f$ (a similar, simplified notation was used in \cite{anile_russo}). 

Assuming that the discontinuity surface is a plane normal to the $x$ axis, we can write components $n^\mu$ as $n^\mu = W_s (V_s,1,0,0)$, where $W_s = 1/\sqrt{1 - V_s^2}$. The quantity $V_s$ has a natural interpretation of the coordinate velocity of the discontinuity. In this case Rankine--Hugoniot conditions have the following algebraic form
\begin{eqnarray}
\label{rankine_hugoniot_1}
\left[ \left[ \rho W^2 - \kappa \rho \right] \right] V_s & = & \left[ \left[ \rho W^2 v^x \right] \right],  \\
\label{rankine_hugoniot_2}
\left[ \left[ \rho W^2 v^x \right] \right] V_s           & = & \left[ \left[ \rho W^2 (v^x)^2 + \kappa \rho \right] \right],  \\
\label{rankine_hugoniot_3}
\left[ \left[ \rho W^2 v^y \right] \right] V_s           & = &  \left[ \left[ \rho W^2 v^x v^y \right] \right],  \\
\label{rankine_hugoniot_4}
\left[ \left[ \rho W^2 v^z \right] \right] V_s           & = &  \left[ \left[ \rho W^2 v^x v^z \right] \right],
\end{eqnarray}
where we have assumed an ultrarelativistic equation of state.

In the case of zero tangential velocity only Eqs.~(\ref{rankine_hugoniot_1})--(\ref{rankine_hugoniot_2}) are relevant. The shock wave velocity can be expressed as
\begin{equation}
V_s = \left[ \left[ \rho W^2 v^x \right] \right] / \left[ \left[ \rho W^2 - \kappa \rho \right] \right],
\end{equation}
which, inserted into Eq.~(\ref{rankine_hugoniot_2}), gives
\begin{equation}
\left[ \left[ \rho W^2 v^x \right] \right]^2 = \left[ \left[ \rho W^2 (v^x)^2 + \kappa \rho  \right] \right]  \left[ \left[ \rho W^2 - \kappa \rho \right] \right].
\end{equation}
The above equation yields
\begin{equation}
\left( \rho / \bar \rho \right)^2 - 2 \left( \Theta + 1 \right) \left( \rho / \bar \rho \right) + 1 = 0,
\end{equation}
where $\Theta = W^2 \bar W^2 (v^x - \bar v^x)^2/(2 \kappa (1 - \kappa))$, and the only physical solution for $\rho$ is given by
\begin{equation}
\rho = \bar \rho \left( 1 + \Theta + \sqrt{(1 + \Theta)^2 - 1} \right).
\end{equation}
This equation, similarly to the rarefaction wave described above, gives the relation between the post-shock density $\rho$ and the post-shock velocity $v^x$.

For a case with non-vanishing tangential velocity a similar calculation can be done. We start by multiplying both sides of equation (\ref{rankine_hugoniot_2}) by $V_s$ and add the result to equation (\ref{rankine_hugoniot_1}).  Then, the expression for $\rho W^2$ can be written as
\begin{equation}
\label{rho_W}
\rho W^2 = \bar \rho \bar W^2 \frac{(\bar v^x - V_s)(1 - \bar v^x V_s)}{(v^x - V_s)(1 - v^x V_s)},
\end{equation}
where we have assumed that $V_s \neq v^x$. Eqs.~(\ref{rankine_hugoniot_3}), (\ref{rankine_hugoniot_4}) give the following expression for the square of $v^t$
\begin{equation}
\label{v_t}
(v^t)^2 = \frac{(V_s - \bar v^x)^2 \bar \rho^2 \bar W^4 (\bar v^t)^2}{\rho^2 W^4 (V_s - v^x)^2}.
\end{equation}
Inserting these two results to Eq.~(\ref{rankine_hugoniot_1}) yields an equation which, after suitable rearrangement of terms, can be written as
\begin{eqnarray}
\lefteqn{ \frac{\bar \rho \bar W^2 (v^x - \bar v^x) V_s}{(1 - \bar v^x V_s)(1 - v^x V_s) (v^x - V_s)} } \\
& & \times \left\{ \left( 1 -\bar v^x V_s \right) \left[ (1 - v^x V_s) (1 - \bar v^x V_s) \right. \right. \nonumber \\
& & \left. \left.- \frac{1}{c^2_s} (v^x - V_s)(\bar v^x - V_s) \right] - (\bar v^t)^2 (1 - v^x V_s) (1 - V_s^2)  \right\} =  0. \nonumber
\end{eqnarray}
Physical values of $V_s$ can be now expressed in terms of $v^x$ as the solutions of the cubic equation
\begin{eqnarray}
\left( 1 -\bar v^x V_s \right) \left[ (1 - v^x V_s)(1 - \bar v^x V_s) - \frac{1}{c^2_s} (v^x - V_s)(\bar v^x - V_s) \right]  & & \\
- (\bar v^t)^2 (1 - v^x V_s) (1 - V_s^2)  & = & 0. \nonumber
\end{eqnarray}
This can be done, for instance, by using one of the Cardano's formulae. Finally, by combining Eqs.~(\ref{v_t}) and (\ref{rho_W}) we can obtain the following expression for the post-shock density $\rho$ as the function of the post-shock velocity $v^x$
\begin{equation}
\label{post_shock_density}
\rho = \frac{\bar \rho \bar W^2 (\bar v^x - V_s)\left[ (1 -(v^x)^2)(1 - \bar v_x V_s)^2 - (\bar v^t)^2 (1 - v^x V_s)^2 \right]}{(v^x - V_s)(1 - v^x V_s)(1 - \bar v^x V_s)}.
\end{equation}

\section{Contact discontinuity}

For $V_s = v^x$ Eqs.~(\ref{rankine_hugoniot_1})--(\ref{rankine_hugoniot_4}) have a non-trivial solution where $\bar v^x = v^x (= V_s)$ and $\bar \rho = \rho$, while velocities $v^y$ and $v^z$ can exhibit an arbitrary jump. This corresponds to the so called contact discontinuity---the one co-moving with the fluid.

Obviously, such a discontinuity can only be present in case of non-vanishing tangential velocities. In this respect, there is a qualitative difference between solutions of the Riemann problem for the ultrarelativistic equations of state and those obtained for the perfect gas equation of state. In the latter case only the pressure and the normal velocity $v^x$ have to be continuous across the contact discontinuity, and there is no such requirement for the baryonic density and the specific internal energy. Thus, in case of the perfect gas equation of state, one usually observes the contact discontinuity also in a strictly one-dimensional problem (with vanishing tangential velocities), and such a discontinuity is absent in analogous solutions with ultra-relativistic equation of state.

\section{Solutions of the Riemann problem}

\begin{figure}[t!]
\begin{center}
\includegraphics[width=80mm]{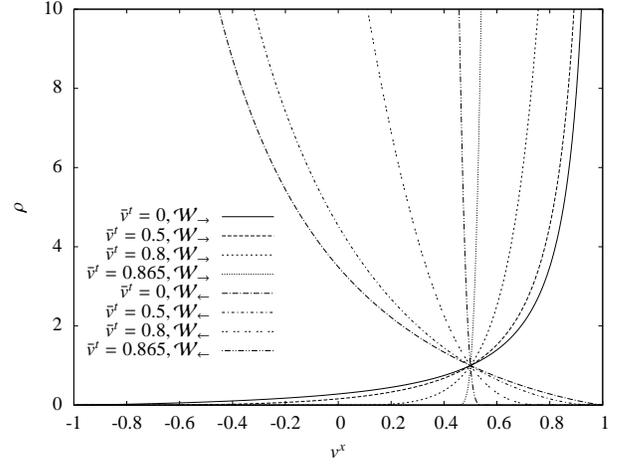}
\vspace{-1em}
\end{center}
\caption{The dependence of the energy density $\rho$ on the velocity $v^x$ behind the wave for the ultra-relativistic equation of state with $c^2_s = 1/3$. Different curves refer to values of the tangential velocity $\bar v^t$ in front of the wave equal to 0, 0.5, 0.8, and 0.865. The velocity $\bar v^x$ in front of the wave was equal 0.5, and the density $\bar \rho$ was set to 1. Increasing curves correspond to the right moving waves, while decreasing ones to the left moving waves.}
\label{fig_rho}
\end{figure}

The distinction between a shock and a rarefaction wave is based on the relation between the pressure in front and behind the wave \cite{taub}. If the pressure $\bar p$ in front of the wave is larger than the pressure $p$ behind it, we are dealing with a rarefaction. The converse case with $p > \bar p$ corresponds to a shock wave. Let $\rho = \mathcal S_{\rightarrow(\leftarrow)}(v^x)$ denote the post-shock energy density $\rho$ understood as a function of the post-shock velocity $v^x$, as it can be computed from Eq.~(\ref{post_shock_density}). As usual, the directions of the arrows correspond to the direction from which the fluid enters the wave. A similar function giving the energy density behind the front of the rarefaction wave will be denoted by $\rho = \mathcal R_{\rightarrow (\leftarrow)} (v^x)$. It follows from results of the preceding sections that the general expression for the energy density behind a wave $\mathcal W_{\rightarrow (\leftarrow)}$ can be written as
\begin{equation}
\rho = \mathcal W_\rightarrow (v^x) =
\left\{ \begin{array}{ll}
\mathcal R_\rightarrow(v^x), & v^x < \bar v^x, \\
\mathcal S_\rightarrow(v^x), & v^x \geq \bar v^x
\end{array} \right.
\end{equation}
for a right moving wave, and
\begin{equation}
\rho = \mathcal W_\leftarrow (v^x) = 
\left\{ \begin{array}{ll}
\mathcal S_\leftarrow(v^x), & v^x < \bar v^x, \\
\mathcal R_\leftarrow(v^x), & v^x \geq \bar v^x
\end{array} \right.
\end{equation}
for a left moving one. Here $\bar v^x$ refer to the velocity in front of the wave. Such functions are illustrated on Fig.~\ref{fig_rho} for different values of the tangential velocity in front of the wave $\bar v^t$.

\begin{figure}[t!]
\begin{center}
\includegraphics[width=80mm]{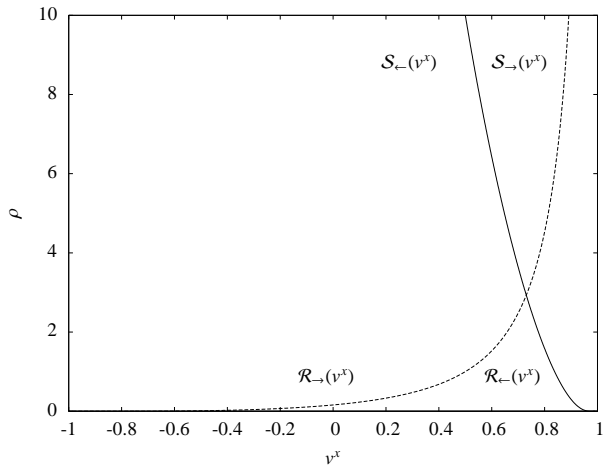}
\vspace{-1em}
\end{center}
\caption{Intersection of the graphs of $\rho(v^x)$ for left and right moving waves. Here both Riemann states correspond to $\bar v^x = 1/2$, $\bar v^t = 1/2$. The energy densities in both states differ: $\rho_L = 10$, $\rho_R = 1$. The curves are computed for the ultra-relativistic equation of state with $c^2_s = 1/3$.}
\label{fig_intersection}
\end{figure}

Given two initial states $L$ and $R$ we can always compute both functions $\rho = \mathcal W_\leftarrow (v^x)$ and $\rho = \mathcal W_\rightarrow (v^x)$, and find the intersection of their graphs. This occurs for some $v^x_\ast$ and $\rho_\ast$ common for both intermediate states $L_\ast$ and $R_\ast$. Such an intersection has been depicted on Fig.~\ref{fig_intersection} for some arbitrary states $L$ and $R$.

In order to complete solving the Riemann problem, one only has to find locations of the interfaces between different states in the solution. The location of the shock wave $\mathcal S$ is given by its speed $V_s$, which can be easily computed after the value of $v_\ast^x$ has been established. The contact discontinuity $\mathcal C$, dividing both states $L_\ast$ and $R_\ast$, travels with the velocity $v_\ast^x$. The velocity of the head of the rarefaction wave is given by the expression for $\xi_\pm$ (plus for $\mathcal R_\rightarrow$, minus for $\mathcal R_\leftarrow$) computed for the suitable Riemann state. The location of the tail of the rarefaction wave can be established by the condition that the velocity $v^x$ in the rarefaction wave should reach the value of $\rho_\ast$. The velocity of the tail is given by $\xi_\pm$ computed for the suitable intermediate state (adjacent to the rarefaction), but the straightforward application of formula (\ref{eigenvalues}) requires a prior calculation of $v^t$ in this state. The values of $v^t$ in both intermediate states can be easily computed from Eq.~(\ref{rho_W}) (for the state behind the shock wave) and from Eq.~(\ref{R_tilde}) (for the state adjacent to the rarefaction wave).

\begin{figure}[t!]
\begin{center}
\includegraphics[width=80mm]{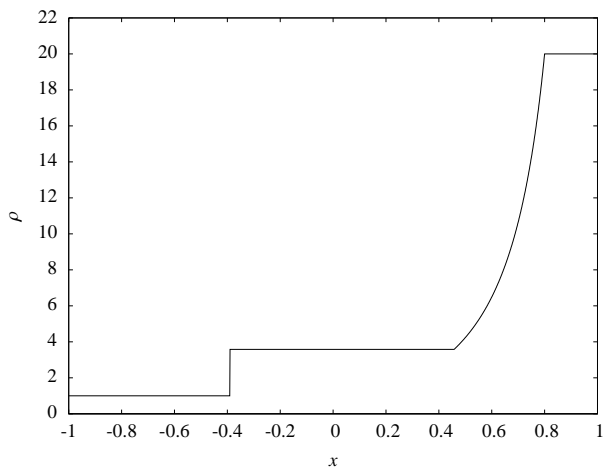}
\vspace{-1em}
\end{center}
\caption{Time snapshot of the solution of the Riemann problem for $t = 1$. The left initial state is given by $\rho_L = 1$, $v^x_L = 1/2$, $v^t_L = 1/3$ and the right state by $\rho_R = 20$, $v^x_R = 1/2$, and $v^t_R = 1/2$. }
\label{fig_solution2}
\end{figure}

\begin{figure}[th!]
\begin{center}
\includegraphics[width=80mm]{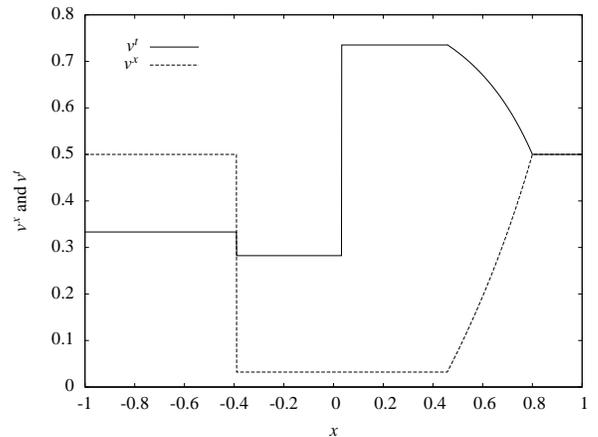}
\vspace{-1em}
\end{center}
\caption{Time snapshot of the solution of the Riemann problem with the same initial data as on Fig.~\ref{fig_solution2}. Here the solid line corresponds to the velocity $v^t$, while the dotted one depicts $v^x$.}
\label{fig_velocities}
\end{figure}

An example of the solution of the Riemann problem for the ultrarelativistic equation of state with $c^2_s=1/3$ is shown on Figs.~\ref{fig_solution2} and \ref{fig_velocities}. Here the left initial state was given by $\rho_L = 1$, $v^x_L = 1/2$, $v^t_L = 1/3$ and the right state by $\rho_R = 20$, $v^x_R = 1/2$, $v^t_R = 1/2$. It is interesting to note the presence of a contact discontinuity in the tangential velocity on Fig.~\ref{fig_velocities}.

\section{Comparison of the results with solutions for the perfect gas equation of state}

Solutions presented in preceding sections can be compared with solutions of the Riemann problem obtained for the perfect gas equation of state in \cite{marti_mueller, pons_marti_mueller}. These are, in general, completely different solutions, however, as pointed out in the second section, the perfect gas equation of state $p = (\gamma - 1)n \epsilon$ tends to $p = (\gamma - 1) \rho$ in the case where $\rho = n + n \epsilon \approx n \epsilon$. Such a condition can be imposed on initial data by assuming that $n \ll n \epsilon$. It can be observed that the solutions of the Riemann problem with the perfect gas equation of state tend to those for the ultrarelativistic equation of state as $n/\epsilon \to 0$, where the solutions are understood as functions of the self-similarity variable $\xi = x/t$.

\begin{figure}[th!]
\begin{center}
\includegraphics[width=80mm]{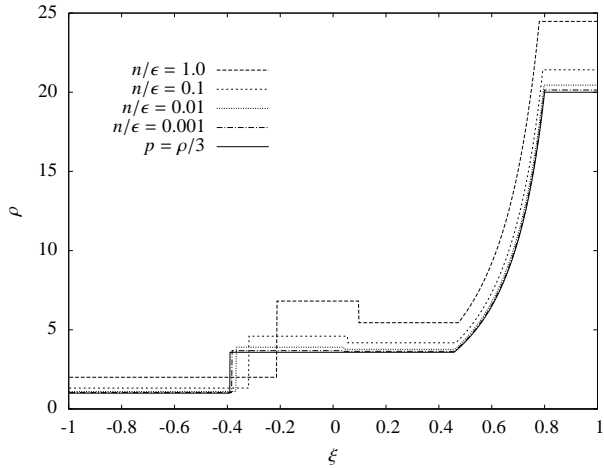}
\vspace{-1em}
\end{center}
\caption{Comparison of solutions obtained for ultrarelativistic (solid line) and perfect gas equations of state (dotted lines). Different solutions for the perfect gas equation of state were obtained for the initial data corresponding to $n/\epsilon$ equal 1.0, 0.1, 0.01, and 0.001 in both Riemann states. Other parameters of the initial states were set to $p_L = 1/3$, $v^x_L = 1/2$, $v^t_L = 1/3$, $p_R = 20/3$, and $v^x_R = v^t_R = 1/2$.}
\label{fig_comparizon}
\end{figure}

It should, however, be noted that although characteristic speeds of propagation of rarefaction waves, shocks, and contact discontinuity tend to those obtained for the ultra-relativistic equation of state, they are different in each of the examined solutions. Thus, having a solution for the perfect gas equation of state which is very close to the one for the ultrarelativistic equation of state in the self-similarity variable $\xi$, it is always possible to consider a sufficient time $t$, after which both solutions, treated as functions of $x$, will vastly differ on an arbitrarily large subset of the domain.

The comparison has been performed for solutions with different values of initial pressures and velocities (both normal and tangential to the initial discontinuity) basing on numerical schemes provided by Mart\'{\i} and M\"{u}ller \cite{living_review} and our solutions. In all examined cases the solutions for perfect gas equation of state tend to those for ultrarelativistic one in a similar way. An example is shown on Fig.~\ref{fig_comparizon}, where we have plotted the energy density of a solution corresponding to the ultrarelativistic equation of state with $c^2_s=1/3$ (solid line) together with densities computed for the perfect gas equation of state with $\gamma = 4/3$ (dotted lines). All solutions of this example were obtained for the following initial conditions $p_L = 1/3$, $v^x_L = 1/2$, $v^t_L = 1/3$, $p_R = 20/3$, and $v^x_R = v^t_R = 1/2$ (the solution the for ultrarelativistic equation of state is thus the same as the one on Figs.~\ref{fig_solution2} and \ref{fig_velocities}). Different solutions for the perfect gas equation of state were computed assuming the values of $n/\epsilon$ in both initial states equal to 1.0, 0.1, 0.01, and 0.001.

\section{Summary}

We have presented an exact solution of the Riemann problem for the ultra-relativistic equation of state, with arbitrary initial velocities, both normal and tangential to the initial discontinuity. Such a solution can be used for testing and construction of the numerical schemes which solve relativistic Euler equations in $(3 + 1)$ dimensions. In fact, our original motivation for dealing with the problem presented in this paper was to provide a test solution for numerical studies of hydrodynamical perturbations in the cosmology of the early universe, where the ultrarelativistic equation of state is a frequent choice.

We also point out that the boosting mechanism described in  \cite{aloy_rezzolla} in the context of astrophysical jets is also exhibited by the solution of this paper. This can be observed on Fig.~\ref{fig_velocities}, where the tangential velocity $v^t$ in the region behind the rarefaction wave is larger than any of the velocities of the initial states.

We have also compared our solution with a similar one obtained for the perfect gas equation of state in \cite{marti_mueller, pons_marti_mueller} in the limit of vanishing baryonic density. In all examined cases solutions for the perfect gas equation of state, treated as functions of the self-similarity variable $\xi$, tend to those for the ultra-relativistic equation of state.

\section*{Acknowledgments}

We would like to thank Bruno Giacomazzo for his useful comments on this paper.


\end{document}